\documentclass[a4paper,12pt]{article}
\usepackage[latin1]{inputenc}
\usepackage{graphicx}
\usepackage{multicol}
\usepackage{float}
\usepackage{color}
\usepackage{fancyhdr}
\usepackage{amsmath,amssymb,latexsym}
\usepackage{stmaryrd}
\usepackage{pstricks}
\usepackage{listings}
\usepackage{subfigure}
\usepackage{xspace}
\usepackage{algorithm, algorithmic}
\usepackage{natbib}
\bibpunct{(}{)}{,}{a}{,}{,}
\title{Using parallel computation to improve Independent Metropolis--Hastings based estimation}
\author{}
\author{P.~Jacob$^{1,2}$, C.P.~Robert$^{1,2}$, M.H.~Smith$^3$\\
$^1$Universit\'e Paris-Dauphine, $^2$CREST, Paris, and\\ $^3$NIWA, Wellington}
\usepackage{etoolbox}

\apptocmd{\thebibliography}{\raggedright}{}{}

\begin{document}

\maketitle

\begin{abstract} 

In this paper, we consider the implications of the fact that parallel raw-power can be exploited by a generic
Metropolis--Hastings algorithm if the proposed values are independent from the current value of the Markov
chain. In particular, we present improvements to the independent Metropolis--Hastings algorithm that
significantly decrease the variance of any estimator derived from the MCMC output, at a null 
computing cost since those improvements are based on a fixed number of target density evaluations that can be
produced in parallel. The techniques developed in this paper do not jeopardize the Markovian convergence
properties of the algorithm, since they are based on the Rao--Blackwell principles of
\cite{gelfand:smith:1990}, already exploited in \cite{casella:robert:1996}, \cite{atchade:perron:2005} and
\cite{douc:robert:2010}. We illustrate those improvements both on a toy normal example and on a classical
probit regression model, but stress the fact that they are applicable in any case where the independent
Metropolis--Hastings is applicable. 

\end{abstract}

\section{Introduction}

The Metropolis--Hastings (MH) algorithm provides an iterative and converging scheme to sample from a complex
target density $\pi$.  Each iteration of the algorithm generates a new value of the Markov chain that relies on
the result of the previous iteration. The underlying Markov principle is well-understood and leads to a generic
convergence principle as described, e.g., in \cite{robert:casella:2004}. However, due to its Markovian nature,
this algorithm is not straightforward to parallelize, which creates difficulties in slower languages like R
\citep{cran}. Nevertheless, the increasing number of parallel cores that are available at a very low cost
drives more and more interest in ``parallel-friendly'' algorithms, that is, in algorithms that can benefit from
the available parallel processing units on standard computers (see. e.g.,
\citet{holmes:doucet:lee:giles:yau:2010}, \citet{Lee2009a}, \citet{Suchard2010}).

Different techniques have already been used to enhance some degree of parallelism in generic
Metropolis--Hastings (MH) algorithms, beside the basic scheme of running $p$ MCMC algorithms independently in
parallel and merging the results.  For instance, a natural entry is to rely on renewal properties of the Markov
chain \citep{mykland:tierney:yu:1995,robert:1995b,hobert:jones:presnel:rosenthal:2002}, waiting for all $p$
chains to exhibit a renewal event and then using the blocks as iid, but the constraint of Markovianity
cannot be removed.  \cite{rosenthal:2000} also points out the difficult issue of accounting for the burn-in
time: while, for a single MCMC run, the burn-in time is essentially negligible, it does create a significant
bias when running parallel chains (unless perfect sampling can be implemented).  \cite{craiu:meng:2005} mix
antithetic coupling and stratification with perfect sampling.  Using a different approach,
\cite{craiu:rosenthal:yang:2009} rely on $p$ parallel chains to build an adaptive MCMC algorithm, considering
in essence that the product of the target densities over the chains is their target, a perspective that
obviously impacts the convergence properties of the multiple chain.  \cite{corander:gyllenger:koski:2006} take
advantage of parallelization to build a non-reversible algorithm that can avoid the scaling effect of specific
neighborhood structures, hence focussing on a very special type of problem.  

A particular family of MH algorithm is the Independent Metropolis--Hastings (IMH) algorithm, where the proposal
distribution (and hence the proposed value) does not depend on the current state of the Markov chain. Due to
this characteristic, this specific algorithm is easier to parallelize and can therefore be considered as a good
building block toward efficient parallel Markov Chain Monte Carlo algorithms, as will be explained in Section
\ref{sec:iMH}. We will focus on cases where the computation of the likelihood function constitutes the major
part of the execution time in the MH algorithm. A most realistic example of this setting is provided in
\cite{wraith-2009-80}, where the model is based on a very complex Fortran program translating the results of
several cosmological experiments, hence highly demanding in computing time. In this model,
\cite{wraith-2009-80} use adaptive importance sampling and massive parallelization, rather than MCMC. 

The fundamental idea in the current paper is that one can take advantage of the parallel abilities of arbitrary
items of computing machinery, from cloud computing to graphical cards (GPU), in the case of the generic IMH
algorithm, producing an output that corresponds to a much improved Monte Carlo approximation machine at the
same computational cost.  The techniques presented here are related with those explained in \cite{perron:1999}
and more closely to those in \citeauthor{atchade:perron:2005} (\citeyear{atchade:perron:2005}, Section 3.1),
since these authors condition upon the order statistic of the values proposed by the IMH algorithm, although in those
earlier papers the links with parallel computation were not established and hence the implementation of the
Rao-Blackwellization scheme became problematic for long chains.

The plan of the paper is as follows: the standard IMH algorithm is recalled in Section \ref{sec:iMH}, followed
by a description of our improving scheme, called here ``block Independent Metropolis--Hastings'' (block IMH).
This improvement depends on a choice of permutations on $\{1,\ldots,p\}$ that is described in details in
Section \ref{sec:permut}. We demonstrate the connections between block IMH and Rao--Blackwellization in Section
\ref{sec:RB}. Results for a toy example are presented throughout the paper and a realistic probit regression
example is described in Section \ref{sec:applications} as an illustration of the method.

\section{Improving the IMH algorithm}\label{sec:iMH}

\subsection{Standard IMH algorithm}\label{sec:iMHb}

We recall here the basic IMH algorithm, assuming the availability of a proposal distribution
that we can sample, and which probability density $\mu$ is known up to a normalization constant. The
independent Metropolis--Hastings algorithm, described in Algorithm \ref{algo:IMH}, generates a Markov chain with
invariant density $\pi$, corresponding to the target distribution.

\begin{algorithm}[H]
\caption{IMH algorithm\label{algo:IMH}}
\begin{algorithmic}[1]
\STATE Set $x_0$ to an arbitrary value
\FOR {$t=1$ to $T$}
	\STATE  Generate $y_t \sim  \mu$
	\STATE\label{basichmacpt}  Compute the ratio:
\[\rho(x_{t-1}, y_t) = \text{min}\left\{
1, \dfrac{\pi(y_t)}{\mu(y_t)}\frac{\mu(x_{t-1})}{\pi(x_{t-1})}
\right\}
\]
	\STATE Set $x_t = y_t$ with probability $\rho(x_{t-1}, y_t)$; otherwise
set $x_t = x_{t-1}$
\ENDFOR
\end{algorithmic}
\end{algorithm}

In the larger picture of Monte Carlo and MCMC algorithms, the IMH algorithm holds a rather special status. It
has certainly been studied more often than other MCMC schemes \citep{robert:casella:2004}, but it is
undoubtedly a less practical solution than the more generic random walk Metropolis--Hastings algorithm. For
instance, it is rather rarely used by itself because it requires the derivation of a tolerably good
approximation to the true target, approximation that most often is unavailable.  On the other hand, first-order
approximations and Metropolis-within-Gibbs schemes are not foreign to calling for IMH local moves based on
Gaussian representations of the targets. The reason theoretical studies of the IMH algorithm abound is that it
has strong links with the non-Markovian simulation methods such as importance sampling. Contrary to random-walk
Metropolis--Hastings schemes, IMH algorithms may enjoy very good convergence properties and may also reach acceptance
probabilities that are close to one. Furthermore, the potentially large gain in variance reduction provided by
the parallelization scheme developped in this paper may counteract the lesser efficiency of the original IMH
compared with the random walk Metropolis--Hastings algorithm.

An important feature of the IMH algorithm, when addressing parallelism, is that it cannot work but in an
iterative manner, since the outcome of step $t$, namely the value $x_t$, is required to compute the acceptance
ratio at step $t+1$. This sequential construction is compulsory for the validation of the algorithm given the
Markov property at its core \citep{robert:casella:2004}.  Nonetheless, given that, in the IMH
algorithm, the proposed values $(y_t)$ are generated independently from the current state
of the Markov chain, $x_t$, it is altogether possible to envision the generation of $T$ proposed values $y_t$
first, along with the computation of the associated ratios $\omega_t = \pi(y_t)/\mu(y_t)$. Once this
computation requirement is completed,
only the acceptance steps need to be considered iteratively. This two-step perspective makes for a huge
saving in computing time when the simulation of the $y_t$'s and the derivation of the $\omega_t$'s can be
achieved in parallel since both the remaining computation of the ratios $\rho(x_{t-1}, y_t)$ given the $\omega_t$'s
and their subsequent comparison with uniform draws typically are orders of magnitude faster.

In this respect the IMH algorithm strongly differs from the random walk Metro\-po\-lis--Has\-tings (RWMH)
algorithm, for which acceptance ratios cannot be processed beforehand because the proposed simulated values depend on
the current value of the Markov chain.  The universal availability of parallel
processing schemes may thus lead to a new surge of popularity for the IMH algorithm. Indeed, when taking advantage
of $p$ parallel processing units, an IMH can be run for $p$ times as many iterations as RWMH, at almost
the same computing cost since RWMH cannot be directly parallelized. 

In order to better describe this increased computing power, we first note that, once $T$ successive values of a
Markov chain have been produced, the sequence is usually processed as a regular Monte Carlo sample to obtain an
approximation of an expectation under the target distribution, $\mathbb{E}_\pi\left[h(X)\right]$ say, for some
arbitrary functions $h$. We propose in this paper a technique that improves the precision of the estimation of
this expectation by taking advantage of parallel processing units without jeopardizing the Markov property.

\begin{figure}[H]
\centering
\includegraphics[width=1\textwidth]{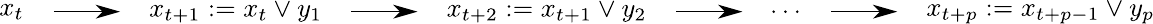}
\caption{\label{fig:IMH}IMH steps between iteration $t$ and iteration $t+p$.}
\end{figure}

Before presenting our improvement scheme, we introduce the notation $\vee$ (read ``or'') for the operator that
represents a single step of the IMH algorithm. Using this notation, given the current value $x_t$ and a sequence of $p$
independent proposed values $y_1, \ldots, y_p \sim \mu$, the IMH algorithm goes from step $t$ to step $t+p$
according to the diagram in Figure  \ref{fig:IMH}.

\subsection{Block IMH algorithm}\label{sec:blockihm}


We propose to take full advantage of the simulated proposed values and of the computation of their
corresponding $\omega$ ratios. 
To this effect, we introduce the {\em block IMH algorithm}, made of successive simulations of blocks of size $p
\times p$. In this alternative scheme, the number of blocks $b$ is such that the number of desired iterations
$T$ is equal to $b * p$, in order to keep the comparison with a standard IMH output fair. Usually $p$ needs not
be calibrated since it represents the number of physical parallel processing units that can be exploited by the
code. However, in principle, this number $p$ can be set arbitrarily high and based on virtual parallel
processing units, the drawback being then an increase in the computing cost. (Note that the block IMH algorithm
can also be implemented with no parallel abilities, still it provides a gain in variance that may counteract
the increase in time.) In the following examples, we take $p$ varying from $4$ to $100$. We first explain how a
block is simulated, and then how to move from one block to the next.

\begin{figure}[H]
\centering
\includegraphics[width=1\textwidth]{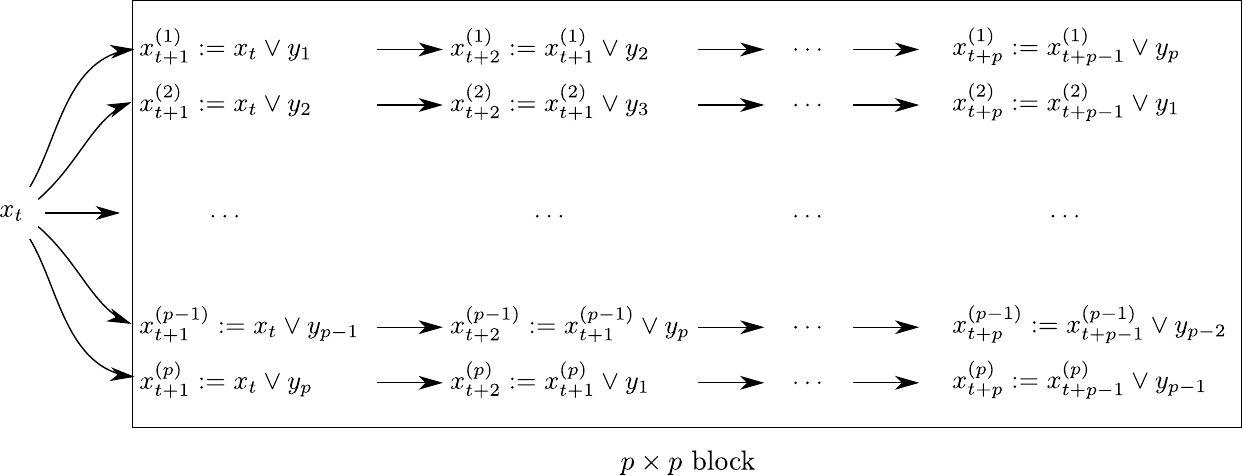}
\caption{\label{fig:IMHblock}Block simulation from step $t+1$ to step $t+p$. Here, circular
permutations of the proposed values are used for illustration purposes.}
\end{figure}

A $p\times p$ block consists in the generation of $p$ parallel generations of $p$ values of Markov chains, all
starting at time $t$ from the current state $x_t$ and all based on the {\em same} proposed simulated values
$y_1, \ldots, y_p$. The different between the $p$ floes is the orders in which those $y_i$'s are included. For
instance, these orders may be the $p$ circular permutations of $y_1, \ldots, y_p$, or they may be instead
random permutations, as discussed in detail (and compared) in Section \ref{sec:permut}. The block IMH algorithm
is illustrated in Figure \ref{fig:IMHblock} for the circular set of permutations.

It should be clear that each of the $p$ parallel chains found in this block is a valid MCMC sequence of length
$p$ when taken separately. As such, it can be processed as a regular MCMC output. In particular, if $x_t$ is
simulated from the stationary distribution, any of the subsequent $x^{(j)}_{t+i}$ is also simulated from the
stationary distribution.  However, the point of the $p$ parallel flows is double:
\begin{itemize}
 \item it aims at integrating out part of the randomness resulting from
the ancillary order in which the $y_k$'s are chosen, getting close to
the conditioning on the order statistics of the $y_k$'s advocated
by \cite{perron:1999};
 \item it also aims at partly integrating out the randomness resulting from
the generation of uniform variables in the selection process, since the
block implementation results in drawing $p^2$ uniform realizations instead
of $p$ uniform realizations for a standard IMH setting. 
\end{itemize}
Both of those points essentially amount to implementing a new Rao--Blackwellization technique (a more precise
connection is drawn in Section \ref{sec:RB}). In an independent setting, each of the $y_k$'s occurs a number
$n_k\ge 0$ of times across the $p$ steps of the $p$ parallel chains, i.e.~for a number $p^2$ of realizations.
Therefore, when considering the standard estimator $\hat\tau_1$ of $\mathbb{E}_\pi\left[h(X)\right]$, based on
a {\em single} MCMC chain,
$$
\hat\tau_1(x_t,y_{1:p}) = \dfrac{1}{p} \sum_{k=1}^p h(x_{t+k})
$$
this estimator necessarily has a larger variance than the double average
$$
\hat\tau_2(x_t,y_{1:p}) = \dfrac{1}{p^2} \sum_{j=1}^p \sum_{k=1}^p h(x^{(j)}_{t+k}) = \dfrac{1}{p^2} \sum_{k=0}^p n_k h(y_{k})
$$
where $y_0 := x_t$ and $n_0$ is the number of times $x_t$ is repeated. (The proof for the reduction of the variance
from $\hat\tau_1$ to $\hat\tau_2$ easily follows from a double integration argument.) We again insist on the
compelling feature that computing $\hat\tau_2$ using $p$ parallel processing units does not cost more time than
computing $\hat\tau_1$ using a single processing unit.

\begin{figure}
\centering
\includegraphics[width=1\textwidth]{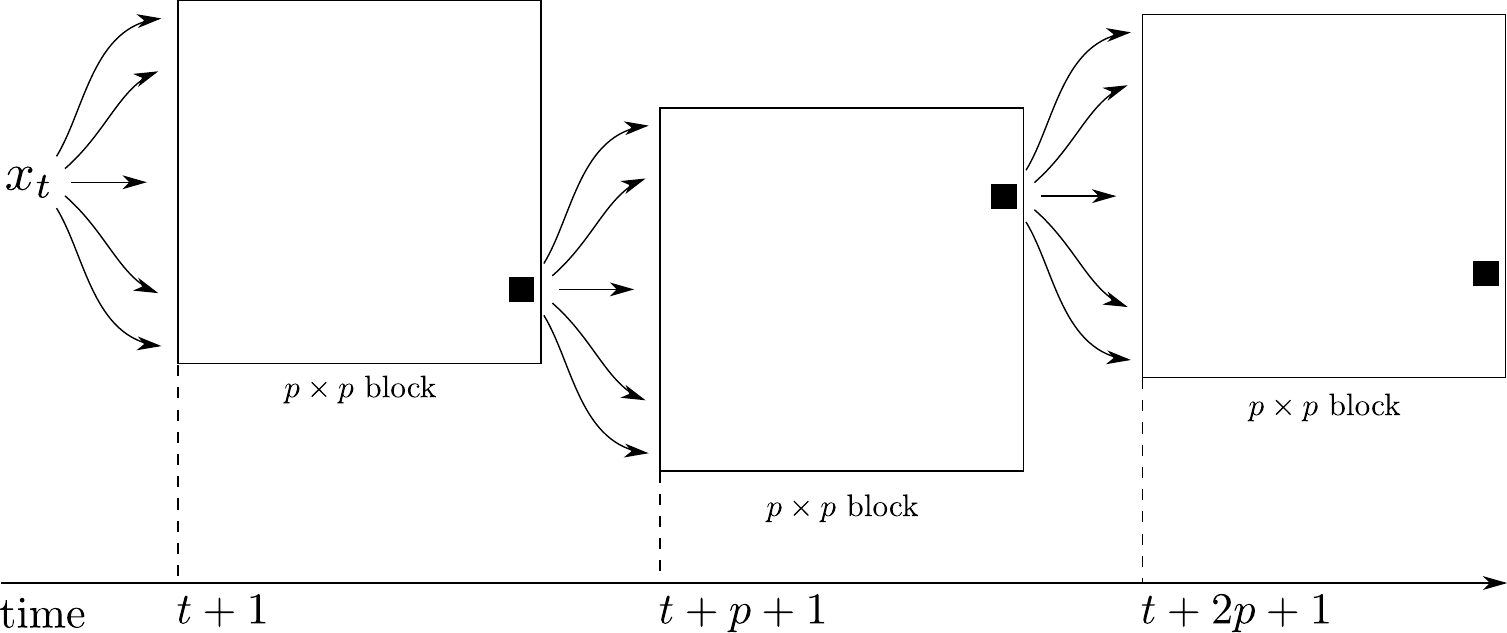}
\caption{\label{fig:IMHnextblock} The block IMH algorithm runs $p$ parallel chains during $p$ steps, then picks one of the 
final values (represented by the black squares) and iterates. (An alternative transition mechanism involves
sampling randomly one of the $p^2$ terms within the block.)}
\end{figure}

In order to preserve its Markov validation, the algorithm must properly continue at time $t+p$. An obvious
choice is to pick one of the $p$ sequences at random and to take the corresponding $x_{t+p}^{(j)}$ as the value
of $x_{t+p}$, starting point of the next parallel block. This mechanism is represented in Figure
\ref{fig:IMHnextblock}. While valid from a Markovian perspective, since the sequences are marginally produced
by a regular IMH algorithm, this means that the chain deduced from the block IMH algorithm is converging at
{\em exactly} the same speed as the original IMH algorithm. An alternative
choice for the starting points of the blocks takes
advantage of the weights $n_k$ on the $y_k$'s that are computed via the block structure.  Indeed, those weights
essentially act as importance weights and they allow for a selection of any of the $p^2$ $x_{t+i}^{(j)}$'s as
the starting point of the incoming block, which corresponds to choosing one of the proposed $y_k$'s with
probability proportional to $n_k$. While this proposal does reduce the length of the resulting chain, it does
not impact the estimation aspects (which still involve all of the $p^2$ values) and it could improve
convergence, given that the weighted $y_k$'s behave like a discretized version of a sample from the target
density $\pi$. We will not cover this alternative any further.

The original version of the block IMH algorithm is described in Algorithm \ref{algo:blockIMH}, 
The algorithm is made of a loop on the $b$ blocks and an inner loop on the $p$ parallel chains of each block. The $p$
steps of this inner loop are actually meant to be implemented in parallel. The output of Algorithm \ref{algo:blockIMH} is double:
\begin{itemize}
\item a standard Markov chain of length $T$, which is 
made of $b$ chains of length $p$, each of which is chosen among $p$ chains 
at line \ref{algo:statepickindex} of Algorithm \ref{algo:blockIMH},
\item a $p \times T$ array $(x_{t}^{k})_{t = 1:T}^{k = 1:p}$, on which the estimator $\hat\tau_2$ is based.
\end{itemize}

\begin{algorithm}
\caption{block IMH algorithm\label{algo:blockIMH}}
\begin{algorithmic}[1]
\STATE Set $x_0$ to an arbitrary value, compute $\omega_0$
\STATE Set $x_\text{start} = x_0$, $\omega_\text{start} = \omega_0$
\STATE Set a block size $p$, and a number of blocks $b$, such that $b * p = T$
\STATE Generate all proposed values $y_1, \ldots, y_T \sim  \mu$
\STATE \label{algo:stateintensive} Compute all ratios $\omega_1, \ldots, \omega_T$
\FOR {$i=1$ to $b$}
  \STATE \label{algo:chooseperm} Choose $p$ permutations $\sigma_1, \ldots,
  \sigma_p$
  \FOR {$k=1$ to $p$}
	  \STATE  Run $p$ steps of an IMH given:
	  \begin{itemize}
	  \item $(x_\text{start}, \omega_\text{start})$
	  \item $p$ proposed values $y_{(i-1)*p + 1}, \ldots, y_{i*p}$ shuffled with 
		    the permutation $\sigma_k$
	  \item the $p$ corresponding ratios $\omega_j$'s
	  \end{itemize}
	  \STATE Save as $x_{(i-1)*p + 1}^{(k)}, \ldots, x_{i*p}^{(k)}$ the resulting chain
  \ENDFOR
\STATE \label{algo:statepickindex} Draw an index $j$ uniformly in $\{1,\ldots, p\}$, set $x_\text{start} = x_{i*p}^{(j)}$, 
set $\omega_\text{start}$ as the corresponding ratio $\omega$.
\ENDFOR
\end{algorithmic}
\end{algorithm}

\subsection{Savings on computing time}

Since the point-wise evaluation of the target density $\pi(y_k)$ is usually the most computer-intensive part of
the algorithm, sampling additional uniform variables has a negligible impact here, as do further costs related
to the storage of vectors larger than in the original IMH. This is particularly compelling since the multiple
chains do not need to be stored further than during a single block execution time.  That is why, although we
sample $p$ times more uniforms in the block IMH algorithm, we still consider it to be running at roughly of the
same cost as the IMH algorithm.  The number of target density evaluations indeed is the same for both and
most often represent the overwhelming part of the computing time in the Metropolis--Hastings algorithm.
Besides, pseudo-random generation of uniforms can also benefit from parallel processing, see e.g.
\cite{LEcuyer01anobjectoriented}.

In the following Monte Carlo experiment, various versions of the block IMH algorithm are compared one
to another, as well as to standard IMH and importance sampling. We stress that
a straightforward reason for not conducting a comparison with a plain parallel
algorithm based on $p$ independent parallel chains is that it does not make
much sense cost-wise.  Indeed, running $p$ parallel MCMC chains of the same
length $T$ does cost $p$ times more in terms  of target density evaluations.
Obviously, if one insists on running $p$ independent chains, for instance as to
initialize an MCMC algorithm from several well-dispersed starting points, each
of those chains can benefit from our stabilizing method, which will improve the
resulting estimation.

The method is presented here for square blocks of dimension $(p, p)$, but blocks could be rectangular as well:
the algorithm is equally valid when using $r \neq p$ permutations, leading to $(r, p)$ blocks.  We focus here
on square blocks because when the machine at hand provides $p$ parallel processing units, then it is most
efficient to simulate the proposed values and the uniforms, and to compute the target densities and the
acceptance ratios at the $p$ proposed values in parallel. 
Once again, the block IMH algorithm with $p\times p$ square blocks has about the same cost as the original
IMH algorithm, because computing target densities and acceptance ratios does more than compensate for the cost
of randomly picking an index at the end of each block.  This amounts to say that line \ref{basichmacpt} of
Algorithm \ref{algo:IMH} and line \ref{algo:stateintensive} of Algorithm \ref{algo:blockIMH} are (by far) the
most computationally demanding ones in the respective algorithms.  

\subsection{Toy example}

We now introduce a toy example that we will follow throughout the paper.
The target $\pi$ is the density of the standard $\mathcal{N}(0,1)$ normal distribution
and the proposal $\mu$ is the density of the $\mathcal{C}(0,1)$ Cauchy distribution. 
Hence, the density ratio is 
\[\omega(x) = \dfrac{\pi(x)}{\mu(x)}
 \propto (1 + x^2) \exp{(-\frac{1}{2} x^2)}\]
We only consider the integral $\int{x\pi(\text{d}x)}$, the mean of $\pi$ equal to zero in this case.
The acceptance rate of the IMH algorithm for this example is around $70\%$. (Note that IMH with higher
acceptance rates are considered to be more efficient, in opposition to other Metropolis--Hastings
algorithms, see \citealp{robert:casella:2004}.)

In all results related to the toy example presented thereafter, $10,000$ independent runs are used to compute
the variance of the estimates. The value of $p$ represents the number of parallel processing units that are
available, ranging from $4$ for a desktop computer to $100$ for a cluster or a graphics processing unit (GPU)
(this value could even be larger for computers equipped with multiple GPUs).  

The results of the simulation experiments are presented in Figures
\ref{fig:barplotpermutations}--\ref{fig:barplotvariousscale} as barplots, which indicate the percentage of
variance decrease associated with the estimators under comparison, the reference estimator being the standard
IMH output $\hat\tau_1$.  In agreement with the block sampling perspective, the same proposed values and uniform
draws were used for all the estimators that are plotted on the same graph (that is, for a given value of $p$), so that the comparison is not
perturbed by an additional noise associated with the simulation.

\section{Permutations\label{sec:permut}}

While the choice of permutations in line \ref{algo:chooseperm} of Algorithm \ref{algo:blockIMH} is irrelevant
for the validation of the parallelization, it has important consequences on the variance improvement and we now
discuss several natural choices.  The idea of testing various orders of the proposed values in a IMH algorithm
appeared in \cite{atchade:perron:2005} where the permutations were chosen to be circular.  We first list
natural types of permutations along with some justifications, and then we empirically compare their impact on
estimation performances for the toy example.

\subsection{Five natural permutations}

Let $\mathcal{S}$ be the set of permutations of $\{1, \ldots, p\}$. Its size is $p!$, 
therefore too large to allow for averaging over all permutations, although this solution would be ideal.  We
consider the simpler option of finding $p$ efficient permutations in $\mathcal{S}$, denoted by $(\sigma_1,
\ldots, \sigma_p)$, the goal being a choice favoring the largest possible decrease in the variance of the
estimator $\hat\tau_2$ defined in Section \ref{sec:iMH}.

\subsubsection{Same order}

The most basic choice is to pick the same permutation on each of the $p$ chains: \[\sigma_1 = \sigma_2 = \ldots
= \sigma_p\] This selection may sound counterproductive, 
but we still obtain a significant decrease in the variance of $\hat\tau_2$ using this set of permutations, when
compared with $\hat\tau_1$. The reason for the improvement is that $p$ times more uniforms are used in
$\hat\tau_2$ than in $\hat\tau_1$, leading to a natural Rao-Blackwellization phenomenon that is studied in
details in Section \ref{sec:RB}.  Nonetheless this simplistic set of permutations is certainly not the best
choice since it does not integrate out the ancillary randomness resulting from the arbitrary ordering of the
proposed values.

\subsubsection{Circular permutations}

Another simple choice is to use circular permutations. For $1\le i \le p$, we define
\[
\sigma_i(1) = i, \sigma_i(2) = i+1, \ldots, \sigma_i(p-i+1) = p, \sigma_i(p-i+2) = 1, \ldots, \sigma_i(p) = i - 1.
\]
An appealing property of the circular permutations is that each simulated value $y_k$ is proposed and evaluated at a different
step for each chain. However, a drawback is that the order is not deeply changed: for instance $y_{k-1}$ will always
be proposed one step before $y_k$ except for one of the $p$ chains, for which $y_k$ is proposed first.

\subsubsection{Random permutations}

A third choice is to use random orders, that is random shufflings of the sequence $\{1, \ldots, p\}$. We can either draw
those random permutations with or without replacement in the set $\mathcal{S}$, but considering the cardinality of the
set $\mathcal{S}$ this does not make a large difference. Indeed, it is unlikely to draw twice the same permutation, except
for very small values of $p$.

\subsubsection{Half-random half-reversed permutations}

A slightly different choice of permutations consists in drawing $p/2$
permutations at random ($p$ is taken to be even here to simplify the notations). Then, denoting the first $p/2$ permutations by
$\sigma_1, \ldots, \sigma_{p/2}$, we define for $1\le k\le p/2$:
\[ \sigma_{k+p/2}(1) = \sigma_{k}(p), \sigma_{k+p/2}(2) = \sigma_{k}(p-1), \ldots
\sigma_{k+p/2}(p) = \sigma_{k}(1)\,.\]
The motivation for this inversion of the orders is that, in the second half of the permutations, the opposition
with the ``reversed'' first half is maximal. This choice, suggestion of Art Owen (personal communication),
aims at minimizing the possible common history among the $p$ parallel chains.
Indeed two chains with the same proposed values in reverse order cannot have a common path of length more than 1.

\subsubsection{Stratified random permutations}

Finally we can try to draw permutations that are far from one another in the set $\mathcal{S}$.  For instance
we can define the lexicographic order on $\mathcal{S}$, draw indices from a low discrepancy sequence on the set
$\{1, \ldots, p!\}$ and select the permutations corresponding to these indices. In a simpler manner, we do use
here a stratified sampling scheme: we first draw a random permutation conditionally on its first element being
$1$, then another permutation beginning with $2$, and so forth until the last permutation which begins with
$p$.

\subsection{A Monte Carlo comparison}

\begin{figure}
\centering
\subfigure{\includegraphics[width=\textwidth]{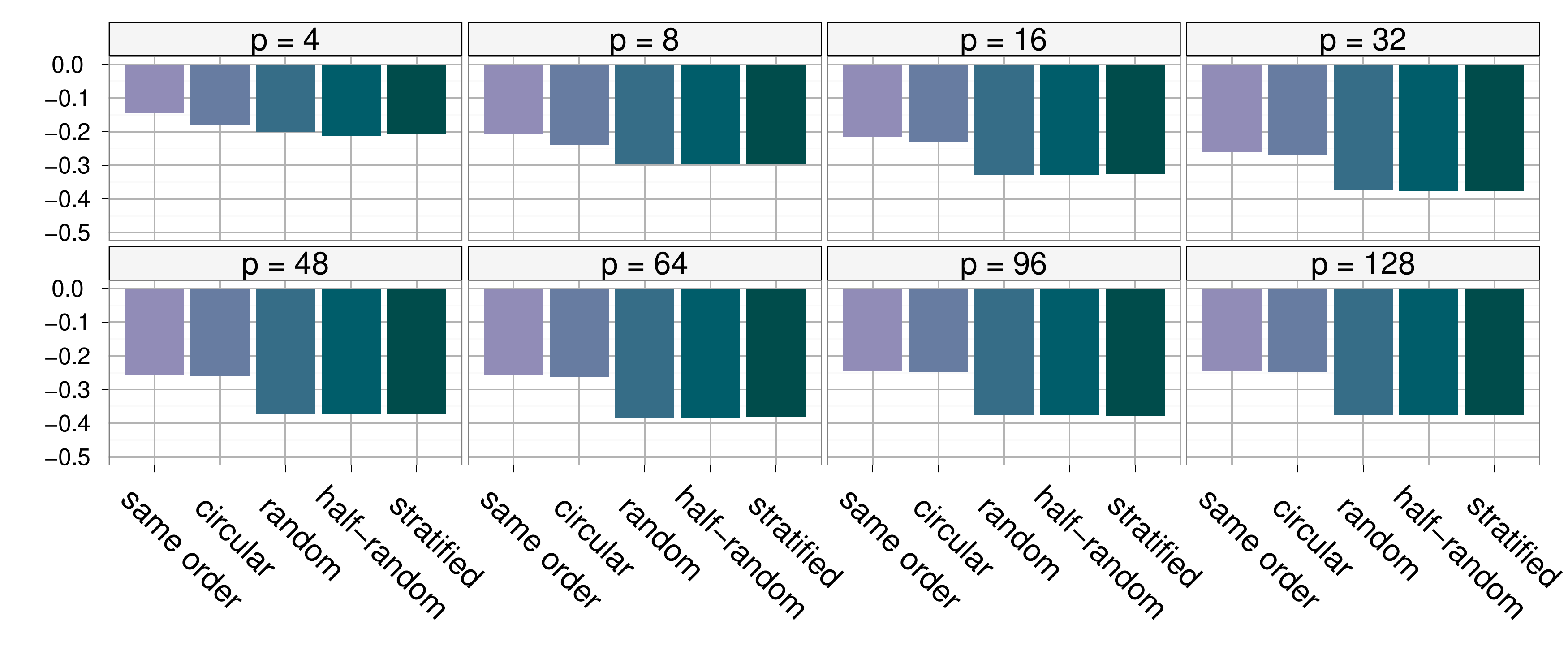}}
\caption{\label{fig:barplotpermutations} Variance reductions, when compared with
the basic estimator $\hat\tau_1$, of the various block estimators $\hat\tau_2$ associated with 
each permutation scheme for several values of $p$.}
\end{figure}

We compare the five described types of permutations on the toy example. Figure \ref{fig:barplotpermutations}
shows the results for various values of $p$, displaying the variance reduction of $\hat\tau_2$ associated with
each of the permutation orders, compared to the variance of the original IMH estimator $\hat\tau_1$. For each
of the $10,000$ independent replications, the block IMH algorithm was launched on one single $p\times p$ block,
e.g. with $b = 1$ using the notation of Section \ref{sec:iMH}, since $b$ plays no role whatsoever in this
comparison.

As mentioned above, using the same order in the $y_k$'s for each of the $p$ parallel chains already produces a
significant decrease of about $20\%$ in the variance of the estimators.  This simulation experiment shows that
the three random permutations (random, half-random half-reversed and stratified) are quite equivalent in terms
of variance improvement and that they are significantly better than the circular permutation proposal, which
only slightly improves over the ``same order'' scheme.  Therefore, in the next Monte Carlo experiments, we will
only use the random order solution, simplest of the random schemes. An amount of improvement like $35\%$ when $p
\geq 32$ is quite impressive when considering  that it is essentially obtained cost-free for a computer with parallel
abilities \citep{holmes:doucet:lee:giles:yau:2010}.

\section{Rao--Blackwellization\label{sec:RB}}

Another generic improvement that can be brought over classical MH algorithms is Rao--Black\-wel\-liza\-tion
\citep{gelfand:smith:1990, casella:robert:1996}.  In this section, two Rao--Black\-wel\-liza\-tion methods are
presented, one that is computationally free and one that, on the contrary, is computationally expensive. We
then implement both solutions within the block IMH algorithm and explain why the ``same order'' scheme already
improves upon the IMH algorithm.

\subsection{Primary Rao--Blackwellization}

Within the standard IMH algorithm of Section \ref{sec:iMHb},  a cost-free improvement can be obtained by a
straightforward Rao--Blackwellization argument.  Given that at step $t+i$, $y_i$ is accepted with probability
$\rho(x_{t+i-1},y_i)$ and rejected with probability $1-\rho(x_{t+i-1},y_i)$, the weight of $y_i$ can be updated by
$\rho(x_{t+i-1},y_i)$ and the weight of the simulated value $y_j$ corresponding to $x_{t+i-1}$ can be similarly updated
by the probability $1-\rho(x_{t+i-1},y_i)$. Considering next the block IMH algorithm, at the beginning of each block we 
can define $p$ weights, denoted by $(w_k)_{k=1}^p$, initialized at $0$ and then, 
for the first of the $p$ parallel chains, denoting by $j$ the index such that $x^{(1)}_{t+i-1}= y_{j}$, we
update these weights at each time $t+i$ as
\begin{align*}
 w_{j} &\leftarrow w_{j} + 1 - \rho(x^{(1)}_{t+i-1},y_i)\\
 w_{i} &\leftarrow w_{i} + \rho(x^{(1)}_{t+i-1},y_i)
\end{align*}
This is obviously repeated for each of the other parallel chains,  ending up with $\sum_k w_k = p^2$. This leads to a new estimator 
$$
\hat\tau_3(x_t,y_{1:p}) = \dfrac{1}{p^2} \sum_{k=0}^p w_k h(y_{k})\,.
$$
This estimator still depends on all uniform generations created within the block, since those weights $w_k$
depend upon the acceptances and rejections of the $y_k$'s made during the block update.  However, along the
steps of the block, the $w_k$'s are basically updated by the expectations of the acceptance indicators
conditionally upon the results of the previous iterations, whereas the $n_k$ of Section \ref{sec:iMH} are
directly updated according to the acceptance indicators. Hence, the $w_k$'s have a smaller variance than the
$n_k$'s by virtue of the Rao--Blackwell theorem, leading to $\hat\tau_3$ necessarily having a smaller variance
than $\hat\tau_2$.

We now discuss a more involved Rao-Blackwellization technique first proposed by \cite{casella:robert:1996}.

\subsection{Block Rao--Blackwellization}\label{sec:iRB2}

Exploiting the Rao--Blackwellization technique of \cite{casella:robert:1996} within each parallel chain
provides via a conditioning argument an even more stable approximation of arbitrary posterior quantities.
As developed in \cite{casella:robert:1996}, for a single Markov chain $(x_1^{(i)}, \ldots, x_p^{(i)})$, 
a Rao--Blackwell weighting scheme on the proposed values $y_t$, with weights $\varphi_t$, is given by a
recursive scheme
$$%
\varphi_t^{(i)} = \delta_t\sum_{j=t}^{p} \xi_{tj} 
$$
where $(t>0)$
$$
\delta_0=1\,,\qquad \xi_{tt}=1\,,\qquad \xi_{tj}=\prod_{u=t+1}^j (1-\rho_{tu})
$$%
and
$$
\delta_t = \sum_{j=0}^{t-1} \delta_j\xi_{j(t-1)}\rho_{jt}\,,
$$
associated with the Metropolis--Hastings ratios
$$
\omega_t = \pi(y_t)/\mu(y_t)\,,\qquad \rho_{tu} = \omega_u/\omega_t \wedge 1\,.
$$
The cumulated computation of the $\delta_t$'s, of the $\rho_{tu}$'s and of the $\xi_{tu}$'s requires an
$\text{O}(p^2)$ computing time. Given that $p$ is usually not very large, this additional cost is not
redhibitory as in the original proposal of \cite{casella:robert:1996} who were considering the application of
this Rao--Blackwellization technique over the whole chain, with a cost of $\text{O}(T^2)$ (see also
\citealp{perron:1999}).

Therefore, starting from the estimator $\hat\tau_2$, the weight $n_k$ counting the number of occurrences of
$y_k$ in the $p\times p$ block can be replaced with the expected number $\varphi_k$ of times $y_k$ occurs in
this block (given the $p$ proposed values), which is the sum of the expected numbers of times $y_k$ occurs in
each of the $p$ parallel chain: \[\varphi_k = \sum_{i = 1}^p \varphi_k^{(i)}\]
Since the $p$ parallel chains incorporate the proposed values with different orders, the $\varphi$'s may differ
for each chain and must therefore be computed $p$ times. Note that the cost is still in $\text{O}(p^2)$ if this
computation can be implemented in parallel. Then, by a Rao-Blackwell argument, $\hat\tau_2$ and $\tau_3$ are dominated 
by $\hat\tau_4$ defined as follows:
$$
\hat\tau_4(x_t,y_{1:p}) = \dfrac{1}{p^2} \sum_{k=0}^p \varphi_k h(y_{k})
$$
Therefore, this Rao--Blackwellization scheme involves {\em no} uniform generation for the
computation of $\hat\tau_4$: the randomness associated with these uniforms is completely integrated out. 

The four estimators defined up to now can be summarized as follows:
\begin{itemize}
\item $\hat\tau_1$ is the basic IMH estimator of $\mathbb{E}_\pi\left[h(X)\right]$,
 \item $\hat\tau_2$ improves $\hat\tau_1$ by averaging over permutations of the proposed values, and by using $p$ times more uniforms than
$\hat\tau_1$,
 \item $\hat\tau_3$ improves upon $\hat\tau_2$ by a basic Rao-Blackwell argument,
 \item $\hat\tau_4$ improves upon the above by a further Rao-Blackwell
argument, integrating out the ancillary uniform variables, but at a cost of $\text{O}(p^2)$.
\end{itemize}
Note that these four estimators all involve the same number $p$ of target density evaluations, which again
represent the overwhelming part of the computing time.

\subsection{A numerical evaluation}

\begin{figure}
\centering
\subfigure{\includegraphics[width=\textwidth]{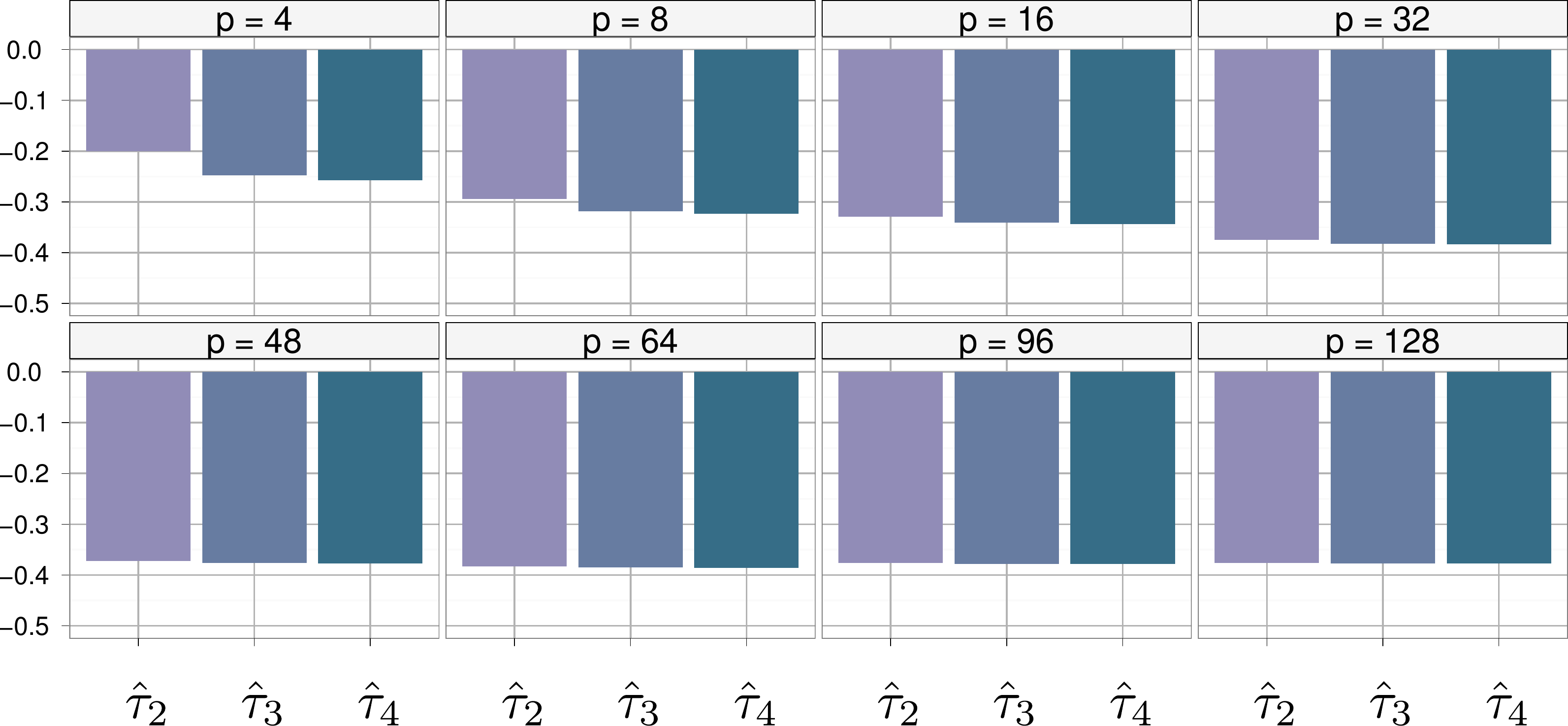}}
\caption{\label{fig:barplotRB} Variance improvement over the basic estimator $\hat\tau_1$ for
three improved block IMH estimators.}
\end{figure}

Figure \ref{fig:barplotRB} gives a comparison between the variances of the three improved estimators defined
above and the variance of the basic IMH estimator. The permutations are random in this case. As was already
apparent on Figure \ref{fig:barplotpermutations}, the block estimator $\hat\tau_2$ is significantly better than
$\hat\tau_1$ for any value of $p$.  Moreover, both Rao-Blackwellization modifications seem to improve only very
slightly the estimation when compared with $\hat\tau_2$, even though the improvement increases with $p$.

Recall that the ``same order'' scheme already provided a significant decrease in the variance of the estimation. In the
light of our results, our interpretation is that using $p$ parallel chains with the same proposed values acts like a
"poor man" Rao--Blackwellization technique since $p$ times more uniforms are used. Specifically, each of the $p$ proposed values is
proposed $p$ times instead of once, thus reducing the impact of each single uniform draw on the overall estimation.

When we use Rao--Blackwellization on top of the block IMH, in the estimators $\hat\tau_3$ and $\hat\tau_4$, we try
indeed to integrate out a randomness that already is partly gone. This explains why, although Rao--Blackwellization
techniques provide a significant improvement over standard IMH, the improvement is much lower and thus rather
unappealing when used in the block IMH setting.  This outcome was at first frustrating since Rao--Blackwellization is
indeed affordable at a cost of only $\text{O}(p^2)$. However, this shows {\em in fine} that the improvement brought by
the block IMH algorithm roughly provides the same improvement as the Rao--Blackwell solution, at a much lower cost.

\subsection{Comparison with Importance Sampling}

The proposal density $\mu$ may also be used to construct directly an importance sampling (IS) 
estimator 
\[ \hat\tau_{IS} = \frac{1}{T} \sum_{t = 1}^T h(y_t) \dfrac{\pi(y_t)}{\mu(y_t)}\,,\] 
where the values $y_t$ are drawn from $\mu$.  It therefore makes sense to compare the IMH
algorithm with an IS approximation because IS is similarly easy to parallelize, and straightforward to program. Furthermore, since the IS estimator does
not involve ancillary uniform variables, it is comparable to the Rao--Blackwellized version of IMH, and hence to the
block IMH. Obviously, IS cannot necessarily be used in the settings when IMH algorithms are used, because the latter are
also considered for approximating simulations from the target density $\pi$. In particular, when considering
Metropolis-within-Gibbs algorithms, IS cannot be used in a straightforward manner, even for approximating integrals.

Before giving numerical results for a comparison run on the toy example, we now explain why in this
comparison we took the number of blocks to be larger than $1$. The previous comparisons
were computed with $b=1$, i.e.~on a single $p \times p$ block and for a large number of
independent runs. The choice of $b$ was then irrelevant since we were comparing methods that were exploiting in
different ways the $p$ proposed values generated in each block. When considering the block IMH algorithm as a
whole, as explained in Section \ref{sec:iMH}, the end of each block sees a new starting value chosen from the
current block. This ensures that the algorithm produces a valid Markov chain.  However, our construction also
implies that the successive blocks produced by the algorithm are correlated, which should lead to lesser
performances than for the IS estimator.

\begin{figure}
\centering
\subfigure{\includegraphics[width=\textwidth]{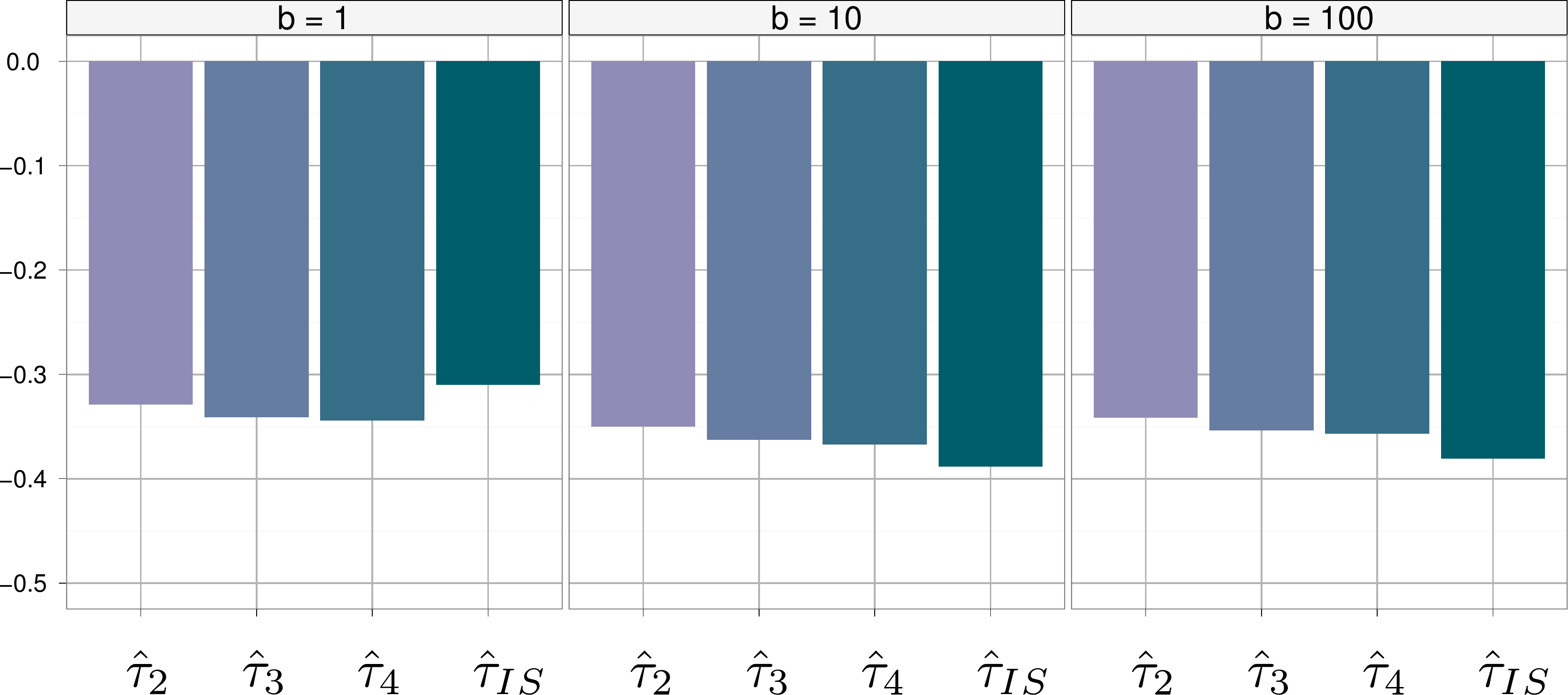}}
\caption{\label{fig:barplotIS} Variance reduction, when compared with
the basic estimator $\hat\tau_1$, of the three improved block IMH estimators, 
and the IS estimator $\hat\tau_{IS}$, for $p=16$ and $b = 1, 10, 100$.}
\end{figure}

In the comparison between IMH and IS, we therefore need to take into account this correlation between successive blocks.
To this effect, we produce the variance reductions for several values of $b$. Those reductions are presented in Figure 
\ref{fig:barplotIS} for $p = 16$ and different values of $b=1, 10, 100$. Once
again, the permutations in the block IMH
algorithm are chosen to be random.

Figure \ref{fig:barplotIS} shows the a priori surprising result that, when selecting $b = 1$ in the experiment,
the variance results are in favor of the block IMH solutions over the IS estimator, but, for any realistic
application, $b$ is (much) larger than $1$. For all $b \ge 10$, the IS estimator has a smaller variance than
the three alternative block IMH estimators, if only by a small margin. (Note that the variance improvement over
the original MCMC estimator is slightly increasing with $b$ despite the correlation between blocks, given that
the correlation between the $p^2$ terms involved in the block IMH estimators is lower than the correlation in
the original MCMC chain.) This experiment thus shows that the block IMH solution gets very close to the IS
estimator, while preserving the Markovian features of the original IMH algorithm.

\section{A probit regression illustration\label{sec:applications}}

In order to evaluate the performances of the parallel processing presented in this paper on a realistic example,
we examine its implementation for the probit model already analyzed in \cite{marin:robert:2010} for the comparison of model
choice techniques because the ``plug-in" normal distribution based on MLE estimates of the first two moments works perfectly
as an independent proposal.

A probit model can be represented as a natural latent variable model
in that, if we consider a sample $z_1,\ldots,z_n$ of $n$ independent latent variables
associated with a standard regression model, i.e.~such that
$z_i|\theta\sim\mathcal{N}\left(x_i^\text{T}\theta,1\right)$,
where the $x_i$'s are $p$-dimensional covariates and $\theta$ is the vector of regression coefficients,
then $y_1,\ldots,y_n$ such that
$$
y_i = \mathbb{I}_{z_i>0}
$$
is a probit sample. Indeed, given $\theta$, the $y_i$'s are independent Bernoulli rv's with
$\mathbb{P}(y_i=1|\theta)=\Phi\left(x_i^\text{T}\theta\right)$ where $\Phi$
is the standard normal cdf. The choice of a prior distribution for the probit model is open to debate,
but the above connection with the latent regression model induced \cite{marin:robert:2007}
to suggest a $g$-prior model, $\theta\sim\mathcal{N}\left(0_p,n(X^\text{T}X)^{-1}\right)$,
with $n$ as the $g$ factor and $X$ as the regressor matrix.

While a Gibbs sampler taking advantage of the latent variable structure is implemented in \cite{marin:robert:2010} and
earlier references \citep{albert:chib:1993b}, a straightforward Metropolis--Hastings algorithm may be used as well,
based on an independent proposal $\mathcal{N}({\hat\theta, c \widehat\Sigma})$, where $\hat\theta$ is the MLE estimator,
$\widehat\Sigma$ its asymptotic variance, and $c$ a scaling factor.

As in \cite{marin:robert:2010} and \cite{girolami:calderhead:2010}, we use the R Pima Indian benchmark dataset
\citep{cran}, which contains medical information about $332$ Pima Indian women with seven covariates and one explained
binary diabetes variable.

\begin{figure}
\centering
\subfigure{\includegraphics[width=\textwidth]{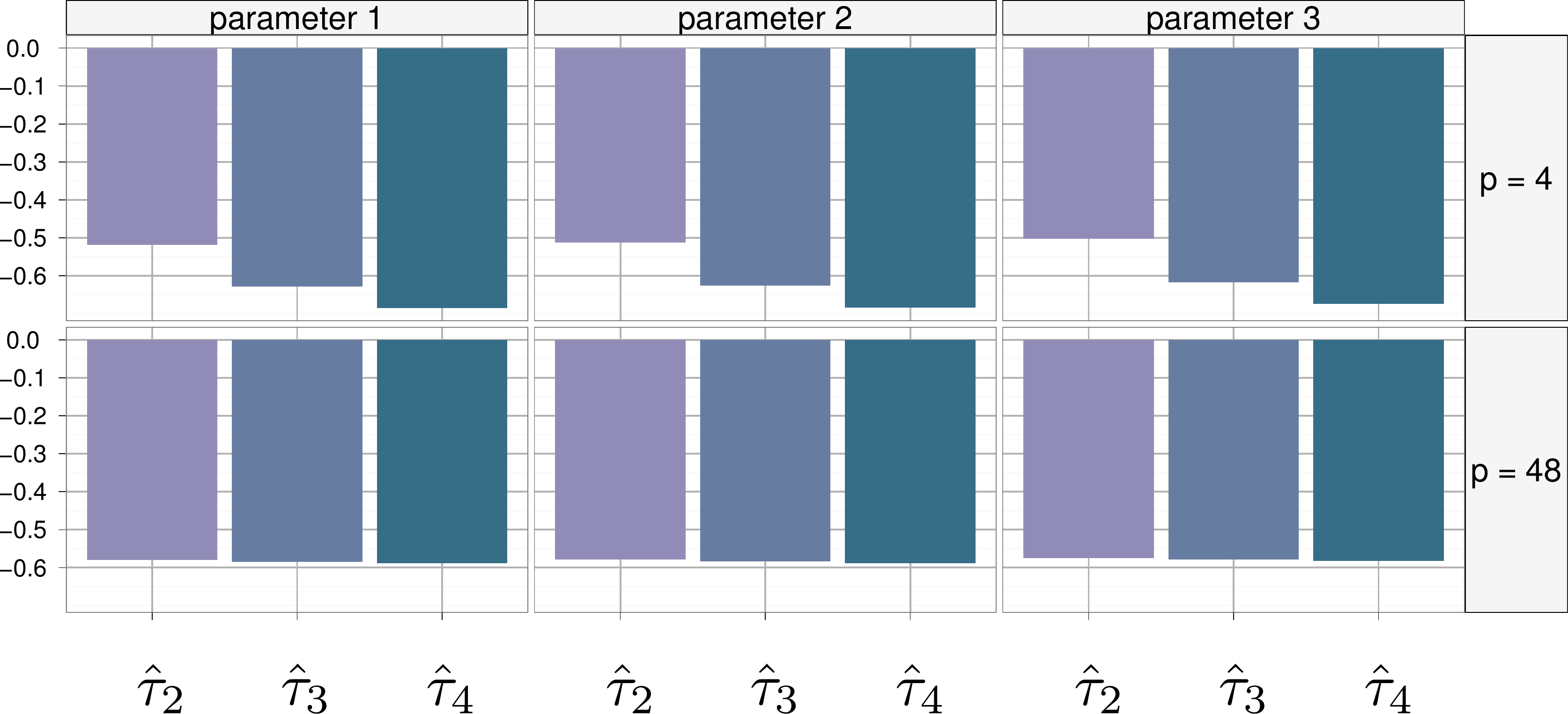}}
\caption{\label{fig:barplotprobit} Variance reduction, when compared with
the basic estimator $\hat\tau_1$, of the three improved block IMH estimators for $p = 10$ and each of the
parameters.}
\end{figure}

For the purpose of illustrating the implementation of the block IMH algorithm, we only consider here three covariates,
namely plasma glucose concentration (with coefficient $\theta_1$), diastolic blood pressure (with coefficient
$\theta_2$) and diabetes pedigree function (with coefficient $\theta_3$). We are interested in the posterior mean of
those three regression parameters. In our experiment, we ran $10,000$ independent replications of our algorithm to
produce a reliable evaluation of the variance of the estimators under
comparison.  In Figure \ref{fig:barplotprobit} we present the variance
comparison of
the four estimators described in Section \ref{sec:RB}, for $p = 4$ and $p =
48$ and for each of the three regression parameters. In the independent
proposal,
the scale factor is chosen to be $3$ since pilot runs showed that it led to an acceptance rate around $37\%$, 
with thus enough rejections to exhibit improvement by Rao--Blackwellization.

The results shown in Figure \ref{fig:barplotprobit} confirm the huge decrease
in variance
previously observed in the toy example.  The gains represented in those figures indicate that the block estimator
$\hat\tau_2$ is nearly as good (in terms of variance improvement) as the Rao--Blackwellized block estimators
$\hat\tau_3$ and $\hat\tau_4$, especially when $p$ moves from $4$ to $48$. This
confirms the previous interpretation
given in Section \ref{sec:RB} that the block IMH algorithm provides a cost-free Rao--Blackwellization as well as a
partial averaging over the order of the proposed values.

The fact that the toy example showed decreases in the variance that were around $35\%$ whereas the probit regression
shows decreases around $60\%$ is worth discussing.  The amount of decrease in the variance differs from
one example to the other, but it is more importantly depending on the acceptance rate of the Metropolis--Hastings
algorithm.  In fact, Rao--Blackwellization and permutations of the proposed values are useless steps if the acceptance
rate is exactly $1$. On the opposite, it may result in a significant improvement when the acceptance rate is low (since
the part of the variance due to the uniform draws would then be much more important).

\begin{figure}
\centering
\subfigure{\includegraphics[width=\textwidth]{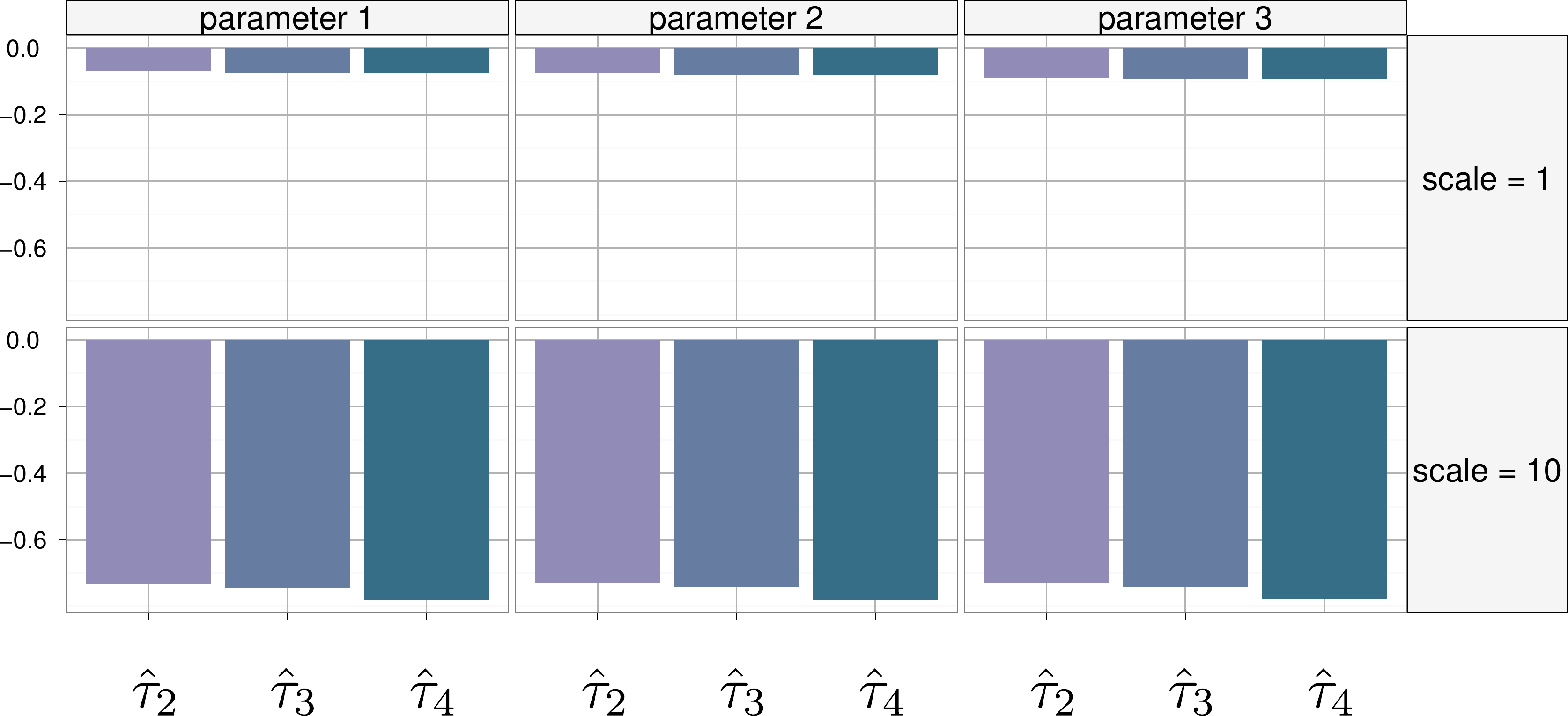}}
\caption{\label{fig:barplotvariousscale} Variance comparison for $p = 16$ and
two scaling factors: $c = 1$, with an
associated acceptance rate of $96\%$ (top) and $c = 10$, with an associated
acceptance rate of $8\%$ (bottom).}
\end{figure}

To illustrate the connection between the observed improvement and the acceptance rate, we propose in Figure
\ref{fig:barplotvariousscale} a variance comparison for $p = 16$ and two
scaling factors $c$ of the proposal covariance matrix in the
probit regression model. 
In the previous experiment, we have used $c = 3$, which leads to an acceptance ratio around $37\%$. Here, if we take $c
= 1$, the acceptance ratio rises to $96\%$, and hence almost all the proposed values are accepted. In this case permuting
the proposed values and using Rao--Blackwellization techniques does not bring much of a variance decrease (Figure
\ref{fig:barplotvariousscale}, top).  On the other hand, if we take $c = 10$,
the acceptance ratio drops down to $8\%$
and the observed decrease in variance is huge. In this second case using all the proposed values gives much better
results than relying on the standard IMH estimator, which is only based on $8\%$ of the proposed values that were
accepted (Figure \ref{fig:barplotvariousscale}, bottom).

The difference observed with this range of scaling factors is thus in agreement with the larger decrease in variance
observed for the probit regression.
This is a positive feature of the block IMH method, since in a complex model, the target distribution is most
often poorly approximated by the proposal and thus the acceptance rate of the IMH algorithm is quite likely to be low.

\section{Conclusion}

The Monte Carlo experiments produced in this paper have shown that the proposed method improves significantly
the precision of the estimation, when compared with the standard IMH algorithm.  Beyond these examples, we see
multiple situations where the block IMH algorithm can be used to improve the estimation in challenging
problems.  First, as already stated, the IMH algorithm can be used in Metropolis-within-Gibbs algorithms
\citep{gilks:best:tan:1995}. Obviously if a single IMH step is performed for each component of the state, then
the block IMH technique cannot be applied without incurring additional costs. However, it is also correct to
update each component multiple times instead of once. Furthermore, an uniform Gibbs scan is rarely the optimal
way to update the components and more sophisticated schemes have been studied, resulting in random scan Gibbs
samplers and adaptive Gibbs samplers \citep{AdaptiveGibbs}, where the probability of updating a given component
depends on the component and is learned along the iterations of the
algorithm. Hence if a component is
updated $n$ times more often than another, $n$ IMH can be performed in a row, which allows the use of the block
IMH technique with $p = n$.

IMH steps are also used within sequential Monte Carlo (SMC) samplers (\citealp{chopin:2002},
\citealp{delmoral:doucet:2003}), to diversify the particles after resampling steps. In this context, an
independent proposal can be designed by fitting a (usually Gaussian) distribution on the particles. If the move
step is repeated multiple times in a row, for instance to ensure a satisfying particle diversification, then
the block IMH algorithm can be used.

Related to SMC, another context where the variance reduction provided by block IMH might be valuable is the
class of particle Markov Chain Monte Carlo methods \citep{andrieu:doucet:holenstein:2010}. For these methods, a
particle filter is computed at each iteration of the MH algorithm to estimate the target density, and hence it
is paramount to make the most out of the expensive computations involved by those estimates. This is thus a
natural framework for parallelization.

As a final message, the block IMH method is close to being $100\%$ parallel (except for the random draw of an
index at the end of each block). Since parallel computing is getting increasingly easy to use, the free
improvement brought by $\hat\tau_3$ is available for all implementations of the IMH algorithm.  Furthermore,
even without considering parallel computing, since the method uses the most of each target density evaluation,
it brings significant improvement when computing the target density is very costly. In such settings, the cost
of drawing $p^2$ instead of $p$ uniform variates is negligible and the block IMH algorithm thus runs in about
the same time as the standard IMH algorithm. We note that the time required to complete a block in the
algorithm is essentially the maximum of the $p$ times required to calculate the density ratios $w_i$.
Therefore, if these times widely vary, there could be a diminishing saving in computation time as $p$ increases
for both the standard IMH and the block IMH algorithms. Nonetheless, even in such extreme cases, using
$\hat\tau_3$ in the block IMH algorithm would bring a significant variance improvement at essentially no
additional cost.

%

\section*{Acknowledgements}

The work of the second author (CPR) was partly supported by the Agence Nationale de la Recherche (ANR, 212, rue de Bercy
75012 Paris) through the 2009 project ANR-08-BLAN-0218 {\sf Big'MC} and the 2009 project ANR-09-BLAN-01 {\sf EMILE}.
Pierre Jacob is supported by a PhD fellowship from the AXA Research Fund.  Since this research was initiated during the
Valencia 9 Bayesian Statistics conference, the paper is dedicated to Jos\'e Miguel Bernardo for the organization of this
series of unique meetings since 1979. Discussions of the second author with participants during a seminar in Stanford
University in August 2010 were quite helpful, in particular the suggestion made by Art Owen to include the ``half-inversed
half-random" permutations. The authors are grateful to Julien Cornebise for helpful discussions, in particular those leading
to the stratified strategy, and to Fran\c cois Perron for his advice on the permutations. Comments and suggestions from the
editorial team of JCGS were most helpful in improving the presentation of our results.


\end{document}